# A Key Encapsulation Mechanism from Low Density Lattice Codes

Reza Hooshmand

Department of Electrical Engineering,

Shahid Sattari Aeronautical University of Science and Technology, Tehran, Iran

Email: rhooshmand@ssau.ac.ir

**Abstract**: Key Encapsulation Mechanisms (KEMs) are a set of cryptographic techniques that are designed to provide symmetric encryption key using asymmetric mechanism (public key). In the current study, we concentrate on design and analysis of key encapsulation mechanism from low density lattice codes (KEM-LDLC) to go down the key size by keeping an acceptable level of security. The security of the proposed KEM-LDLC relies on the difficulty of solving the closest vector problem (CVP) and the shortest basis problem (SBP) of the lattices. Furthermore, this paper discusses other performance analyses results such as key size, error performance, and computational complexity, as well as conventional security analysis against applied attacks. Reducing the key size is performed by two approaches: (i) saving the generation sequence of the latin square LDLC's parity-check matrix of as a part of the secret key set; (ii) using the hermite normal form (HNF) of the latin square LDLC's generator matrix as part of the public key set. These enhancements enable us to attain greater efficiency and security compared to earlier code-based KEMs.

## 1. Introduction

With the advancement of quantum computers expected in the in the foreseeable future, widely employed public main cryptosystems, like RSA [1] and ElGamal [2] will be easily broken [3]. Research into cryptographic algorithms capable of withstanding quantum computer-based attacks is a main concentration of dynamic studies. Post-quantum cryptosystems, however, are designed to remain secure even when faced with quantum computer-based attacks [4]. Cryptographic primitives from codes and lattices are some of the most encouraging options of post-quantum cryptography and their security is according to the durability of widely recognized problems concerning the coding theory and lattices, respectively. As a matter of fact, code-based and lattice-based cryptosystems serve as effective alternatives to the traditional cryptosystems that rely on number theory. Moreover, code-based and lattice-based cryptosystems has been raised in the post-quantum cryptography (PQC) standardization process of the national institute of standards and technology (NIST) [5]. During this standardized process, numerous candidates for public-key encryption (PKE), digital signatures, and key encapsulation mechanisms (KEM) are introduced. In fact, the development of effective and resilient post-quantum KEMS is crucial and pressing research objective, as indicated by the recently NIST request for post-quantum cryptographic systems [6-10]. As mentioned earlier, these efforts are aimed at addressing the growing need for secure cryptographic solutions in respect to upcoming threats of quantum computing. A KEM encompasses a set of algorithms that make two parties able to securely set a shared secret key over a public communication channel under particular conditions. Once the shared symmetric key is established through the KEM, it can be utilized in symmetric cryptographic algorithms to perform essential security functions, such as encryption and authentication. A KEM is a set of three algorithms (KeyGen, Encaps, Decaps) elaborated as follows: (i) KeyGen($\lambda$) is a key generation algorithm that takes the target security level $\lambda$ as input and returns a couple (SK, PK) of secret and public keys; (iii) Encaps(PK) is an encapsulation algorithm that takes the public key PK as input. It first generates a symmetric key $\mathcal{K}$ then computes its encapsulated value $c$; (iv) Decaps (SK, $c$) recovers a symmetric key $\mathcal{K}$ from an encapsulated value $c$. It receives as input an encapsulated value $c$ and a secret key SK. It backs either a symmetric key $\mathcal{K}$ or the failure symbol $\perp$ [11].

As the landscape of cybersecurity keeps on to evolve, understanding and implementing these advanced coding techniques will be crucial for developing next-generation code-based encryption standards capable of withstanding both current and future threats. Polar codes [12, 13] and low-density lattice codes (LDLCs) [14, 15] are categories of linear codes that can attain the channel capacity by using efficient encoding and decoding. Hence, due to the capabilities of polar codes and LDLCs, it is argumentative to use such codes in the code-based cryptography. These codes are exploited in distinct cryptographic schemes such as in public key cryptosystems [16-19], physical layer encryption schemes [20, 21], secure channel coding schemes [22, 23], secret key cryptosystem [24], identification scheme [25], key encapsulation mechanism [26] and hybrid encryption scheme [27]. LDLCs are explained by a sparse parity check matrix and their design and decoding shares similarities with LDPC codes since they utilize belief propagation (BP) decoding on sparse graphs. One promising approach relies in the realm of code-based cryptography, particularly through the use of LDLCs. These codes leverage the mathematical structures of lattices, which manifest resilience against quantum attacks, making them ideal candidates for secure key encapsulation mechanisms. By concentrating on LDLCs, which are characterized by their efficient encoding and decoding processes [28], researchers aim to create robust encryption methods that not only boost security but also reinforce performance in practical applications. In developing a code-based key encapsulation mechanism utilizing LDLCs, this research seeks to address the dual challenges of security and efficiency in cryptographic protocols. This mechanism facilitates secure key exchange, which is essential for ensuring confidentiality in digital communications.

In this study, the researcher presents a key encapsulation mechanism based on low density lattice code (KEM-LDLC) that utilizes the practical features of codes and lattices at the same time for its security and efficiency compared to lattice-based encryption systems. However, its security depends on the difficulty of solving the closed vector problem (CVP) and the shortest basis problem (SBP) of the lattices [29, 30]. In fact, in the proposed KEM-LDLC, we have utilized the basics of code-based and lattice-based at the same time, which can be set relationship between code-based and lattice-based cryptosystems. It can be said that the proposed KEM-LDLC leverages the unique properties of LDLCs to make a robust framework for secure key exchange. The carefully structured algorithms, including KEM-LDLC.KeyGen, KEM-LDLC.Encaps, and KEM-LDLC.Decaps, make easy the generation and sharing of symmetric keys between legitimate parties while ensuring resilience against potential security threats. By employing a hermite normal form (HNF) [31, 32] and addressing the complexities of lattice-based cryptography, KEM-LDLC not only enhances security but also provides efficient encoding and decoding processes, making it appropriate for modern communication systems.

Looking forward, the KEM-LDLC stands to assist significantly to the field of post-quantum cryptography, offering a reliable alternative in a landscape growingly concerned with the implications of quantum computing on traditional cryptographic systems. As further research explores the optimization of its parameters and practical implementations, the KEM-LDLC can pave the way for more secure and efficient key exchange protocols. By combining this mechanism into various applications, we can foster a future where secure communications are not only feasible but also resilient against evolving threats in the digital landscape. By integrating the strengths of LDLCs, the proposed system aims to obtain a balance between computational feasibility and cryptographic robustness.

The framework of the current study is organized in the following way. In Section 2, we provide a concise overview of LDLCs. Subsequently, in Section 3, we describe the proposed KEM derived from LDLCs. Sections 4 and 5 present our investigation of the KEM-LDLC, focusing on security and efficiency analyses, respectively. Finally, we summarize our findings in Section 6.

## 2. Low Density Lattice Codes

An $n$-dimensional lattice $L$ in $\mathbb{R}^n$ is the arrangement of all integral integrations of $n$ linearly independent vectors $g_1, g_2, \cdots, g_n$ in $\mathbb{R}^n$, i.e., $L = \mathcal{L}(g_1, g_2, \cdots, g_n) = \{\sum_{i=1}^n m_i g_i : m_i \in \mathbb{Z}\}$. The vectors $g_i = (g_{i1}, g_{i2}, \cdots, g_{in})$, $i = 1, 2, \cdots, n$ are named *basis vectors* of $L$ and the set of basis vectors $\{g_1, g_2, \cdots, g_n\}$ is called as a *basis* of $L$. A matrix $G = \begin{bmatrix} g_1 \\ \vdots \\ g_n \end{bmatrix} \in \mathbb{R}^{n \times n}$ which has the basis vectors $g_i$, $i = 1, 2, \cdots, n$, as its rows is named *basis matrix* (or *generator matrix*) of $L$. An $n$-dimensional lattice in $\mathbb{R}^n$ is written as $L = \mathcal{L}(G) = \{mG : m \in \mathbb{Z}^n\}$, where each *lattice point* is computed as $x = mG$, $m \in \mathbb{Z}^n$. The lattice $L$ is named as a rational lattice if the generator matrix $G \in \mathbb{Q}^{n \times n}$ is a rational matrix. In this case, there is an $n \times n$ unimodular matrix $U$ such that the product matrix $G' = \text{HNF}(G) = UG$ corresponds to the HNF of $G$. The entries of the HNF $G' = [g'_{ij}]$ possess several key properties: (1) $g'_{ij} = 0$, $\forall j < i$; (2) $g'_{ii} > 0$, $\forall i$; (3) $g'_{ij} \leq 0$ and $|g'_{ij}| < g'_{ii}$, $\forall j > i$ [29].

In this context, the generator matrix of the employed LDLC and its HNF represent two distinct underpinning of the identical lattice, denoted as $\mathcal{L}(G') = \mathcal{L}(G)$. The HNF $G'$ is determined by the lattice $\mathcal{L}(G')$ and does not depend on the specific structure of the generator matrix $G$; thus, no data is given about $G$ itself. Given a lattice $\mathcal{L}$ and a desired vector $x \in \mathbb{R}^n$ (which is usually not part of the lattice), CVP seeks to discover the lattice vector $w \in \mathcal{L}$ that minimizes the distance to $x$, satisfying the condition $\|x - w\| \leq \|x - v\|$, $v \in \mathcal{L}$. This problem is recognized as NP-hard. Another challenging problem associated with lattices is SBP. According to a basis $G'$ for a lattice $\mathcal{L}$ in $\mathbb{R}^n$, the objective is to find the shortest basis $G$, which is defined as the basis with least orthogonality discrepancy, denoted as $\text{odf}(G) = |\det(G^{-1})| \prod_i \|g_i\|$, where $\|g_i\|$ represents the Euclidean norm of the $i$-th row $G$. Currently, a polynomial-time algorithm has not been discovered to efficiently solve the SBP [29, 30].

Concerning an AWGN channel subject to power limitations, an $n$-dimensional lattice code in $\mathbb{R}^n$ is formed by the overlap between $\mathcal{L} \subset \mathbb{R}^n$ and a specified shaping region $B \subset \mathbb{R}^n$. In this scenario, the codewords consist of all lattice points within $B$. This approach makes sure that the strength of the codewords remains appropriately constrained, preventing any values from becoming excessively large. Additionally, the number of lattice points is limited, allowing an attacker to potentially locate the nearest lattice point to the encapsulated value through comprehensive search methods. Furthermore, the implementation of the shaping algorithm adds complexity to the encoding process [33]. Alternatively, an AWGN channel without power constraints imposes no restrictions to power and the efficiency of lattice coding schemes is analyzed as if they were unbounded lattices. In this context, the boundless power AWGN channel serves as a valuable theoretical framework for examining the coding properties of lattices without the consideration of a shaping region. The effectiveness of lattice decoding in a limitless power AWGN channel is assessed using Poltyrev's bound [34]. Poltyrev's bound indicates that it is possible to achieve minimally low error probability in transmission over an limitless power AWGN channel using an $n$-dimensional lattice $L$ if and only if the condition $\sigma^2 < \sqrt[n]{|\det(G)|^2}/(2\pi e)$, where $\sigma^2$ is the noise dispersion [35, 36].

LDLCs are innovative and functional lattice codes that are able to effectively approach channel capacity while also allowing for efficient encoding and decoding. An $n$-dimensional LDLC in $\mathbb{R}^n$ is an $n$-dimensional lattice code in $\mathbb{R}^n$ with a nonsingular generator matrix $G_{n \times n}$, i.e., $|\det(G)| = 1$, where the equality check matrix $H_{n \times n}$ is mandated to be sparse structure. The codewords appear in the format $x = mG$, where $G_{n \times n} = H^{-1}$ is an generator matrix of an $n$-dimensional LDLC and $m \in \mathbb{Z}^n$ is an $n$-dimensional integer-valued vector. The sparsity characteristics of the parity check matrix $H$ is leveraged to create an efficient decoding algorithm for LDLCs, resulting in strong error performance. LDLCs are characterized in line with values of the non-zero entries within the matrix $H$. Degree associated with the rows $r_i$, $i = 1,2,\cdots,n$ refers to the count of non-zero entries in the $i$-th row of the parity check matrix $H$. Similarly, the degree associated with the columns $c_i$ for $i = 1, 2, \cdots, n$ represents the quantity of non-zero entries in $i$-th column of $H$. An $n$-dimensional LDLC is referred to as routinized if the dimensions of line and pillar of its parity check matrix $H$ are all uniformly identical to a shared level $d$. A latin square LDLC is a type of $n$-dimensional standard LDLC with a uniform level $d$, meaning that each row and column of its parity check matrix $H$ contains exactly $d$ non-zero values, differing only in random signs and potential rearrangement. The ordered sequence of these $d$ nonzero values $h_1 \geq h_2 \geq \cdots \geq h_d \geq 0$ is referred to as the sequence that develops the latin square LDLC [37].

In sum, LDLCs exemplify a powerful approach to coding theory that merges efficient encoding and decoding processes with the ability to approach channel capacity. The unique structural properties of LDLCs, mainly their sparsity and the features of the parity check matrix, enable the development of robust decoding algorithms that significantly boost error performance. As communication systems continue to evolve and demand higher data rates and reliability, LDLCs present a compelling solution, particularly in scenarios where traditional methods struggle with complexity and performance limitations. Looking ahead, the exploration of LDLCs over unconstrained power AWGN channels opens new avenues for research and application in modern communication networks. By leveraging Poltyrev's bounds and addressing the challenges posed by infinite lattice point distributions, future work can concentrate on optimizing these codes for practical implementations. As researchers seek to refine decoding techniques and enhance performance in real world conditions, LDLCs stand at the forefront of innovations in error correction and data transmission, promising to play a pivotal role in the next generation of communication technologies.

## 3. The Proposed LDLC-based Key Encapsulation Mechanism

In this segment, we first elaborate the constructions and algorithms of the proposed KEM-LDLC and then make its security and efficiency analyses. Figure 1 illustrates the overall structure of the intended KEM-LDLC. As shown in this figure, the proposed KEM-LDLC compasses three algorithms: (i) KEM-LDLC key generation algorithm, named as KEM-LDLC.KeyGen; (ii) KEM-LDLC encapsulation algorithm, called as KEM-LDLC.Encaps; and (iii) KEM-LDLC decapsulation algorithm, named as KEM-LDLC.Decaps. In fact, the aim of KEM-LDLC is to share the symmetric key $\mathcal{K}$ between valid partners. In this case, the KEM-LDLC.KeyGen takes the security parameter $\lambda$ as input and develops an ephemeral public encapsulation key (PK) and an ephemeral secret decapsulation key (SK) as its outputs. The KEM-LDLC.Encaps, takes PK as input and returns the symmetric key $\mathcal{K}$ and encapsulated value $c$, i.e. the encapsulation of $\mathcal{K}$ by means of a key derivation function (KDF). Then, the encapsulated value $c$ is delivered via a non-secure channel. At the receiver side, given the SK and the encapsulated value $c$, the symmetric key $\mathcal{K}$ can be achieved by executing the KEM-LDLC.Decaps. Otherwise, the decapsulation failure ⊥ is its output.

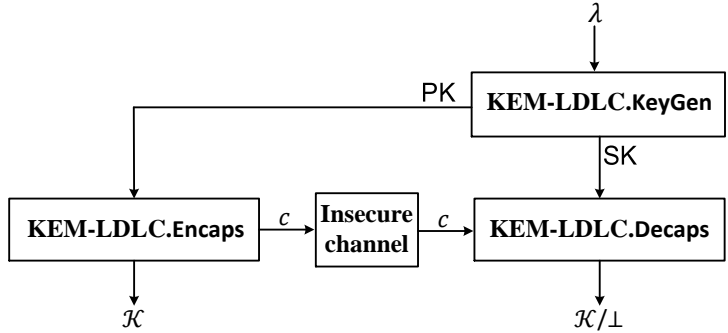

**Figure1.** The overall structure of the proposed KEM-LDLC.

*3.1. KEM-LDLC Key Generation Algorithm (KEM-LDLC.KeyGen)*

The KEM-LDLC.KeyGen is explained as Algorithm 1 in which the target security level $\lambda$ is considered as its input and the SK and PK are considered as its outputs. In this algorithm, the generator matrix $G$ and its HNF $G'$ are two different bases from the same lattice, i.e. $\mathcal{L}(G') = \mathcal{L}(G)$. However, the HNF $G'$ relies on the lattice $\mathcal{L}(G')$ and does not provide any information about the generator matrix $G$. Indeed, retrieving the generator matrix $G$ from its HNF $G'$ is equal to solving the SBP.

<u>Algorithm 1: KEM-LDLC.KeyGen</u>

Input: the target security level $\lambda$;

Output: PK and SK;

1. Given the the target security level $\lambda$, fix the LDLC code parameters $n$, $k$ and $d$.
2. Construct the generating sequence set $\mathcal{H} = \{h_1, h_2, \cdots, h_d\}$ with rational values $0 \leq h_i \leq 1, 1 \leq i \leq d$ such that $h_{i-1} \geq h_i$.
3. Generate the permutation set $\mathcal{P} = \{p_1, p_2, \cdots, p_d\}$ with $d$ distinct integers $1 \leq p_i \leq n$, $1 \leq i \leq d$.
4. Initiate the parity-check matrix $H_{n \times n}$ of an $n$-dimensional latin square LDLC with degree $d$ by using the set $\{\mathcal{H}, \mathcal{P}\}$ such that the determinant of $H$ is equal to one, i.e. $|det(H)| = 1$.
5. Compute the generator matrix of an $n$-dimensional latin square LDLC with degree $d$ as $G_{n \times n} = H^{-1}$.

6. Choose the HNF of generator matrix $G$ as $G'_{n \times n} = \text{HNF}(G) = UG$, where $U$ is an $n \times n$ unimodular matrix.

7. Return SK= $\{\mathcal{H}, \mathcal{P}\}$ and PK= $G'_{n \times n}$.

### 3.2. *KEM-LDLC Encapsulation Algorithm (KEM-LDLC.Encaps)*

In Algorithm 2, the KEM-LDLC.Encaps is intended in which the symmetric key $\mathcal{K}$ is acquired via a KDF. A leveraged KDF:$\{0,1\}^n \to \{0,1\}^{l_{\mathcal{K}}}$ in the KEM-LDLC.Encaps is a digest operation with a variable output length $l_{\mathcal{K}}$, as an instance, Keccak-256 with 32 byte output.

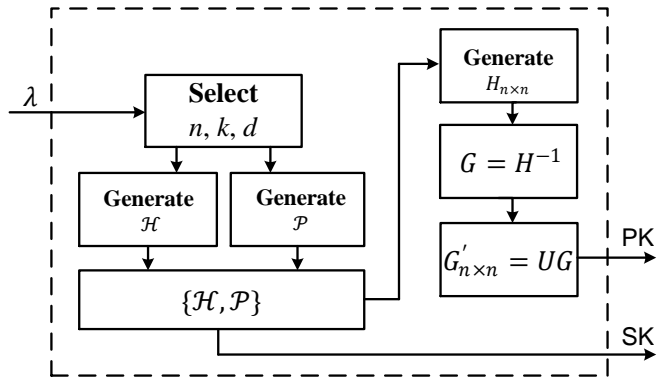

**Figure 2.** The block diagram of the KEM-LDLC.KeyGen.

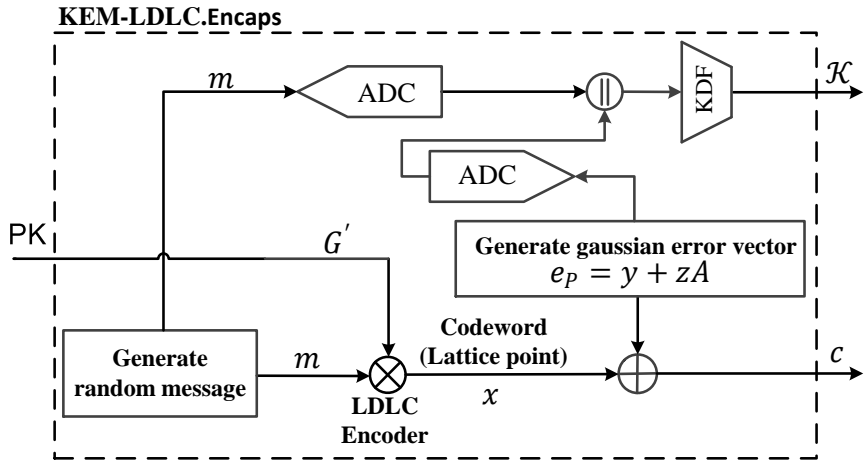

**Figure 3.** The block diagram of the KEM-LDLC.Encaps.

Algorithm 2: KEM-LDLC.Encaps

Data: $n$, $k$, and $d$ parameters of the used LDLC;

Input: A momentary public encapsulation key PK= $G' = \text{HNF}(G)$;

Output: A symmetric key $\mathcal{K}$ and its encapsulated value $c$;

1. Generate a Gaussian perturbation (error) vector $e = (e_1, e_2, \cdots, e_n)$ as $e = y + zA$, where $y = (y_1, y_2, \cdots, y_n) \in \mathbb{R}^n$ is an intentional perturbation (error) vector, $z = (z_1, z_2, \cdots, z_k)$ is a vector of standard normal random variables and $A$ is an arbitrary $k \times n$ matrix. The Gaussian perturbation (error) vector $e \sim \mathcal{N}(\mu, \sigma^2)$ is obtained with mean $E(e) = \mu$ and variance $\sigma^2 = Var(e) = A^T A < 1/(2\pi e)$. In this case, we have LDLCs without any shaping area.
2. Construct an $n$ symbol message $m$ as a set of random integers using the pseudorandom number generator algorithm.
3. Set $\mathcal{K} = \text{KDF}(m \parallel e, l_\mathcal{K})$.
4. Compute the encapsulated value $c = mG' + e_P = m'G + e_P = x + e_P$, where $m' = mU$ and $x = mG' = m'G$ are the lattice points of used LDLC.
5. Return the symmetric key $\mathcal{K}$ and its encapsulated value $c$.

Finally, the transmitter sends the encapsulated value $c$ via the unprotected noiseless channel. Due to the addition of the Gaussian perturbation (error) vector $e \sim \mathcal{N}(\mu, \sigma^2)$ to the codeword $x$ and also by not considering the shaping area in the KEM-LDLC, sending the encapsulated value $c$ through a noiseles channel is equivalent to sending codeword $x$ is limitless power AWGN channel. For such channel, an $n$-dimensional LDLC is large enough that the lattice point (codeword) $x$ can be decoded with the possibility of an arbitrary small error if and only if the Gaussian perturbation (error) vector $e$ has a variance of less than $\sigma^2_{max} = 1/(2\pi e)$. As a matter of fact, the security of the intended KEM-LDLC is according to the idea that by having the public key $G'$, generating the codeword $x$ and obtaining a lattice point closest to $x$ (encapsulated value $c$) by adding the Gaussian perturbation (error) vector $e \sim \mathcal{N}(\mu, \sigma^2)$ with variance $\sigma^2 < 1/(2\pi e)$ to $x$. Clearly, it is impossible to convert the encapsulated value $c$ to the codeword $x$ without knowing SK.

*3.3. KEM-LDLC Decapsulation Algorithm (KEM-LDLC.Decaps)*

In this segment, we elaborate the construction and performance description for the decapsulation algorithm of the proposed KEM-LDLC step by step and in detail (see Fig. 4).

Algorithm 3: KEM-LDLC.Decaps

Data: $n$, $k$, and $d$ parameters of the used LDLC;

Input: An encapsulated value $c$ and SK;

Output: A symmetric key $\mathcal{K}$ or a decapsulation failure $\perp$;

The authorized receiver receives the encapsulated value $c$ and generates the parity-check matrix $H_{n\times n}$ of the utilized LDLC by knowing the symmetric key $\mathcal{K} = \{\mathcal{H}, \mathcal{P}\}$. The parameters required for the proposed KEM-LDLC includes the parity-check matrix $H_{n\times n}$ of employed latin square LDLC and the perturbation (error) vector $e_P = ep$ which is developed from SK as follows:

1. First, retrieve the parity-check matrix $H_{n\times n}$ to calculate $G_{n\times n} = H^{-1}$ and the matrix $P_{d\times n}$ to generate the perturbation (error) vector $e_P$.

2. To build $e_P$, first block the $d$-dimensional integer vector $e = (p_1, p_2, \ldots, p_d)$, which consists of $d$ integer entries of the set $\mathcal{P} = \{p_1, p_2, \ldots, p_d\}$. Then, generate the perturbation (error) vector $e_P = (e_{P_1}, e_{P_2}, \ldots, e_{P_n})$ by multiplying $e$ by the matrix $P_{d\times n}$, i.e., $e_P = eP$.

3. Obtain the generator matrix $G_{n\times n} = H^{-1}$ of the latin square LDLC and the nearest lattice point to the encapsulated value $c$, i.e., estimating the $\hat{x} = \widehat{m'}G$ [6]. The estimation of lattice point $\hat{x}$ is not obtained directly in the LDLC decoding. Rather, the probability density function (pdf) associated with the codeword $x = (x_1, x_2, \cdots, x_n)$, i.e., $\hat{f}_{x|c}(x|c)$, is estimated. In fact, instead of estimating $n$-dimensional pdf of $x$, the $n$ one-dimensional marginal pdfs for each of the components of $x = (x_1, x_2, \cdots, x_n)$, i.e., $\hat{f}_{x_i|c}(x_i|c), i = 1, 2, \cdots, n$, is estimated. Thus, the measured codeword $\hat{x} = (\hat{x}_1, \hat{x}_2, \cdots, \hat{x}_n)$ is obtained.

4. Obtain the estimation of $m' = mU$ as $\widehat{m'} = \lceil \hat{x}G^{-1} \rfloor = \lceil \hat{x}H \rfloor$ where $\lceil \hat{x}H \rfloor$ is the closest integer to $\hat{x}H$. Also, estimate the random message as $\hat{m} = \widehat{m'}U^{-1}$.

5. Retrieve the symmetric key as $\mathcal{K} = \text{KDF}(\hat{m} \parallel e_P, l_{\mathcal{K}})$.

6. Return $\mathcal{K}$ or $\perp$

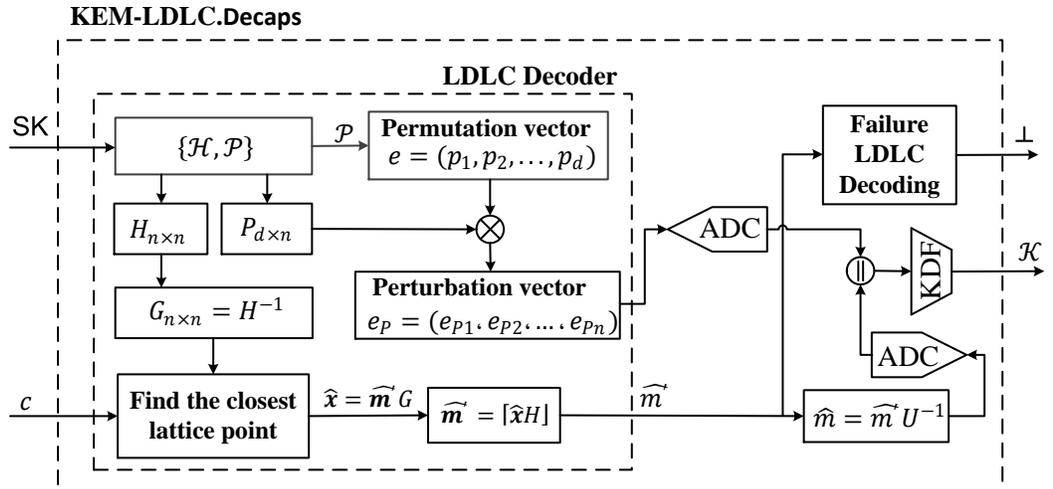

**Figure 4.** The block diagram of the KEM-LDLC.Decaps.

## 4. Efficiency Analysis

In this regard, we assess the intended KEM-LDLC in terms of its error execution, key size, and analytical complexity to assess its overall efficiency.

*4.1. Error performance*

As a result of combining the codeword $x$ with the Gaussian perturbation vector $e \sim \mathcal{N}(\mu, \sigma^2)$, and ignoring any shaping region, transmitting the encapsulated value $c$ over a noiseless channel becomes analogous to sending the codeword $x$ over a limitless power AWGN channel. In the KEM-LDLC, the inputs and outputs of KEM-LDLC.Decaps are integer-valued or rational-valued vectors. Consequently, the error performance is evaluated by the symbol error rate (SER) in relation to the signal-to-noise ratio (SNR). Figure 5 reflects the error execution of the KEM-LDLC with degree $d = 7$ and codewords $n = 100, 1000, 10000$ symbols and generating sequence set $\mathcal{H} = \{0.433, 0.315, 0.196, 0.136, 0.085, 0.076, 0.057\}$. In this figure, the error performance is measured in terms of SER relative to SNR. For example, at the SER of $10^{-5}$, the SNR for the $n = 100, 1000, 10000$ are 3.7, 1.5 and 0.8 dB, respectively. It is obvious that the latin square LDLC with larger codeword lengths have far better error performance than LDLC with smaller codeword lengths. In fact, it has been shown experimentally that for the codeword length $n \geq 1000$ with degree $d \leq 10$, the latin square LDLC parity-check matrix will be such that it has most of the features needed to have the proper LDLC decoding. Therefore, in the KEM-LDLC, the use of LDLC with the codeword length of $n \geq 1000$ and degree $d \leq 10$ will be more efficient.

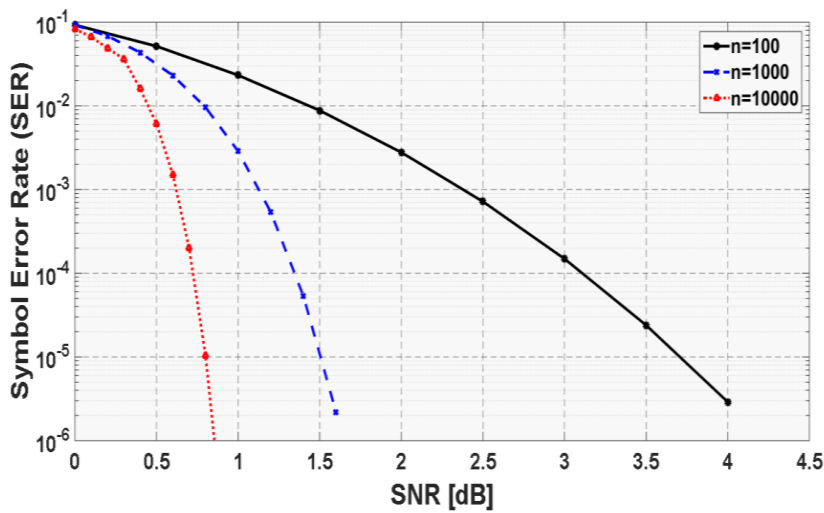

**Figure 5.** SER of the proposed KEM-LDLC with codewords length 100, 1000 and 10000 symbols.

*4.2. Key Size*

In this section, we assess the size of $\mathcal{K} = \{\mathcal{H}, \mathcal{P}\}$ as follows:

1. For $\mathcal{H} = \{h_1, h_2, \cdots, h_d\}$, if the maximum memory needed to store $h_i$, $1 \leq i \leq d$ is $r$ bits, then the memory needed to store $\mathcal{H}$ can be calculated as $\mathcal{M}_\mathcal{H} \leq rd$. For the larger value of $r$, the generating sequence $\mathcal{H}$ becomes more accurate. In the proposed KEM-LDLC, the maximum memory required to store $h_i$, $1 \leq i \leq d$ is set at $r = 16$.

2. For a latin square LDLC of length $n$, the largest probable index, which is $n$, can be one of the entries in the set $\mathcal{P} = \{p_1, p_2, \cdots, p_d\}$. This necessitates $q = \lfloor \log_2^{(n)} \rfloor + 1$ bits for storage. Therefore, the memory needed to store $\mathcal{P}$ is given by $\mathcal{M}_\mathcal{P} \leq qd = (\lfloor \log_2^{(n)} \rfloor + 1)d$. Consequently, the key size can be calculated as $\mathcal{M}_\mathcal{K} = \mathcal{M}_\mathcal{H} + \mathcal{M}_\mathcal{P} \leq \left(r + \lfloor \log_2^{(n)} \rfloor + 1\right)d$.

As an example, for latin square LDLC with a dimension of $n = 10^4$, $d = 7$, $q = 14$ and $r = 16$, the memory requirements can be calculated as $\mathcal{M}_\mathcal{H} \leq 112$ bits and $\mathcal{M}_\mathcal{P} \leq 98$ bits. Thus, the total size of the secret key will be $\mathcal{M}_{SK} = \mathcal{M}_\mathcal{H} + \mathcal{M}_\mathcal{P} \leq 210$ bits. Table 1 presents a comparison of the KEM-LDLC scheme with KEM-PC [26], classic McEliece [6], BIKE [7], and HQC [8] regarding memory usage and communication capacity. The results indicate that the secret key sizes for the KEM-LDLC scheme are significantly smaller than those of other schemes. However, for the same code length and code rate, the public key and encapsulated value sizes of the KEM-LDLC scheme are larger than those of the other schemes.

*4.3. Computational Complexity*

The computational complexity of KEM-LDLC includes two segments: (i) the KEM-LDLC.Encaps complexity and (ii) the KEM-LDLC.Decaps complexity. The computational complexities of such two parts are based on the calculations related to these structures. The complexity of KEM-LDLC.Encaps is declared as $C_{Encaps} = C_{LDLC-Encode}(m) + C_{Add}(e_P) + C_{KDF}(m \parallel e_P)$, where $C_{LDLC-Encode}(m) = \mathcal{O}(n^2 m)$ is the encoding complexity [14] associated with the employed LDLC is determined by the maximum memory needed to store each entry of the rational generator matrix $G$ in binary format, denoted as $m$. In addition, the number of needed binary procedures for addition of $n$-dimensional rational perturbation vector $e_P \sim \mathcal{N}(\mu, \sigma^2)$ to the codeword $x$ is obtained as $C_{Add}(e_P) = \mathcal{O}(nq)$, where $q$ is the maximum needed memory to store each entry of $e_P$ in binary form. Also, $C_{KDF}(m \parallel e_P) = \mathcal{O}(n + n + l_\mathcal{K})$ is the computational complexity of KDF with the input of concatenated $m$ and the perturbation vector $e_P$.

**Table 1.** Comparison between the KEM-LDLC, KEM-PC, classic McEliece, BIKE and HQC schemes.

| Scheme | $(n, k)$ | SK size (kbytes) | PK size (kbytes) | ct size (bytes) |
|---|---|---|---|---|
| KEM-LDLC I | $n = 1000, d = 7$ | $\leq 0.022$ | 61 | 750 |
| KEM-LDLC II | $n = 2000, d = 7$ | $\leq 0.023$ | 244.1 | 1500 |
| KEM-LDLC III | $n = 5000, d = 7$ | $\leq 0.025$ | 1525.85 | 3750 |
| KEM-LDLC IV | $n = 10000, d = 7$ | $\leq 0.026$ | 6103.5 | 7500 |
| KEM-PC I [26] | $n = 1024, k = 816$ | 0.253 | 20.718 | 128 |
| KEM-PC II [26] | $n = 2048, k = 1632$ | 0.559 | 82.875 | 256 |
| KEM-PC III [26] | $n = 4096, k = 3264$ | 1.219 | 331.5 | 512 |
| KEM-PC IV [26] | $n = 8192, k = 6528$ | 2.64 | 1326 | 1024 |
| Mceliece348864 [6] | $n = 3488, k = 2790$ | 6.3 | 255 | 128 |
| Mceliece460896 [6] | $n = 4608, k = 3686$ | 13.25 | 511.875 | 188 |
| Mceliece6960119 [6] | $n = 6960, k = 5568$ | 13.58 | 1022.77 | 226 |
| Mceliece6960119 [6] | $n = 8192, k = 6554$ | 13.75 | 1326 | 240 |
| BIKE-128 [7] | $n = 12323, k = 256$ | 0.274 | 1.5 | 1572 |
| BIKE-128 [7] | $n = 40973, k = 256$ | 0.566 | 5 | 5154 |
| HQC-128 [8] | $n = 23869, k = 256$ | 0.313 | 2.953 | 6017 |
| HQC-256 [8] | $n = 69259, k = 256$ | 0.313 | 8.494 | 17379 |

The calculative complexity of KEM-LDLC.Decaps is declared as $C_{Decaps} = C_{LDLC-Decode}(c) + C_{mul}(\widehat{m'}U^{-1}) + C_{KDF}(\hat{m} \| e_P)$, where $C_{LDLC-Decode}(c) = \mathcal{O}(n)$ is the complexity of LDLC decoding. Also, $C_{mul}(\widehat{m'}U^{-1}) = \mathcal{O}(n^2 l)$ is the number of needed binary procedures to conduct the output of the inferred vector $\widehat{m'}$ by the opposite of unimodular matrix $U$, where $l$ is the maximum memory necessary to store each individual entry of $U$ in binary form. Also, $C_{KDF}(\hat{m} \| e_P) = \mathcal{O}(n + n + l_{\mathcal{K}})$ is the computational complexity of KDF with the input of concatenated $\hat{m}$ and the perturbation vector $e_P$.

## 5. Security Analysis

The security of KEM-LDLC is according to the hardness of well-known lattice problems like the CVP and the SBP. In the proposed KEM-LDLC, finding the generator matrix $G$ of the used LDLC from the public key $G' = \text{HNF}(G)$ requires solving SBP instances of the lattice $\mathcal{L}(G) = \mathcal{L}(G')$. Also, finding the closest lattice vector to the encapsulated value $c$, i.e., $x = m'G$ such that $\|c - x\|$ is minimized, requires solving some CVP instances of the lattice $\mathcal{L}(G) = \mathcal{L}(G')$. In the continuation of this part, we analyze the stability of the intended KEM-LDLC against the numerous assaults that have been applied to the GGH cryptosystem.

*5.1. The Brute-Force Attack*

The brute-force attack is a straightforward yet powerful method employed to compromise cryptographic systems by systematically attempting all possible keys until the correct one is found. In the KEM-LDLC, an attacker systematically evaluates every potential $SK = \{\mathcal{H}, \mathcal{P}\}$ until the accurate one is identified. In the KEM-LDLC, to analyze the number of developing sequence set $\mathcal{H} = \{h_1, h_2, \cdots, h_d\}$, we examine two scenarios: (a) the rational indications of the set $H$, where its entries are rational numbers ranging between 0 and 1; and (b) the binary form of the set $H$, in which its elements are stored as $r$-bit numbers. The quantity of rational representations of $H$ is non-countable, as there are infinitely many rational numbers within the interval from 0 to 1. Consequently, it is not feasible for an attacker to identify the rational representation of $H$ within a polynomial time. The count of binary representations of $H$ is influenced by the parameters $r$ and $d$ and is calculated using the formula $\mathcal{N}_\mathcal{H} = \prod_{i=0}^{d-1}(2^r - i)$. By choosing the parameters $r$ and $d$ appropriately, $\mathcal{N}_\mathcal{H}$ can attain a sufficiently large value. For example, in the proposed KEM-LDLC, setting $r = 16$ and $d = 7$ yields $\mathcal{N}_\mathcal{H} = \prod_{i=0}^{6}(2^{16} - i) \approx 2^{112}$. Also, the number of set $\mathcal{P} = \{p_1, p_2, \cdots, p_d\}$ is computed as $\mathcal{N}_\mathcal{P} = \prod_{i=0}^{d-1}(n - i)$. As an example, if we consider $n = 10^4$ and $d = 7$, we have $\mathcal{N}_\mathcal{P} \approx 2^{93}$. It is crucial for the attacker to identify both $H$ and $P$ in order to successfully create the parity check matrix $H$ of the utilized LDLC.

*5.2. The Ciphertext Only Attack*

In the ciphertext-only attack (COA), the attacker gains access to the encrypted value $c$ through methods like eavesdropping or surveillance, aiming to retrieve the integer-valued vector $m$ from the encapsulated value $c$. In the KEM-LDLC, to effectively carry out such an attack, the attacker must decode a general LDLC without any knowledge of its generator or parity check matrix, which is equivalent to solve the CVP. It has been demonstrated that if the lattice aspect is sufficiently large, the CVP becomes NP-hard. Consequently, by utilizing appropriate parameters for the latin square LDLC, such as $n = 10^4$, the COA is impractical for the proposed KEM-LDLC. Furthermore, given that the security of the KEM-LDLC relies on NP-hard problem, specifically the CVP, KEM-LDLC is resistant to attacks carried out by quantum computers [4].

## 5.3. The Adaptive Chosen Ciphertext Attack

An adaptive chosen ciphertext attack (CCA2) is a sophisticated cryptographic attack where an adversary can select specific encapsulated values for decryption and use the corresponding plaintext outputs to gain information about other encapsulated values. In the KEM-LDLC, the encapsulated value is obtained as $c = mG' + e_P$, where $m$ is a message and $e_P$ is an perturbation vector. In this scenario, if the attacker encrypts one of the two messages $m_1$ and $m_2$, and gets an encapsulated value $c$, then he can signify $m_i$ as a plaintext if $\|m_i G' - c\| < \|m_j G' - c\|$. In the CCA2, given an encapsulated value $c = mG' + e_P$ and for some $m'$, if the attacker inputs $c + m'G'$ to the decryption oracle, the output developed by the decryption oracle is $\hat{m}$. With this regard, the attacker can discover the authentic message $m$ by measuring $m = \hat{m} - m'$. It means that the proposed KEM-LDLC does not meet the criteria for indistinctive characteristics and is vulnerable to CCA2. To ensure security against CCA2, we can implement the Fujisaki-Okamoto modification scheme [38, 39] as follows.

Let $\mathcal{E}_\mathcal{K}$ and $\mathcal{D}_\mathcal{K}$ be a conventional encryption method and a decryption process using a key $\mathcal{K}$, respectively. Let $h$ and $g$ be random oracles. Also, let $c_1 = m'G' + e_P$ and $c_2 = \mathcal{E}_{g(e_P)}(m)$, where $m' = h(m, e_P)$. In this case, the encapsulated value is obtained as $c = c_1 \| c_2 = c = (m'G' + e_P) \| \mathcal{E}_{g(e_P)}(m)$. For the KEM-LDLC.Decaps, by employing the LDLC decoding, first $\widehat{m''} = \widehat{m'}G'$ is estimated and $\widehat{m'} = \widehat{m''}(G')^{-1}$ is obtained. Then, the estimate of perturbation (error) vector $\widehat{e_P} = c_1 - \widehat{m'}G'$ is computed. Now, a random plaintext $\hat{m} = \mathcal{D}_{g(e_P)}(c_2)$ with the secret key $g(e_P)$ is recovered. Finally, if $\left(h(\hat{m}, \hat{e})\right)G' + \widehat{e_P} = c_1$, then the decapsulated information from the oracle is $\hat{m}$. Otherwise, the decapsulation fails. Using the Fujisaki-Okamoto scheme, the adaptive intended ciphertext attack can be stopped for the intended KEM-LDLC.

## 5.4. The Message-Resend Attack

In the message-resend attack [40, 41], the attacker influences the encapsulated value $c$ and his objective is to recover the perturbation (error) vector $e_P$. Let us consider that Alice encrypts a random message $m$ as $c_1 = mG' + e_{P_1}$ and delivers it to Bob. In contrast, Oscar affects the encapsulated value $c_1$. In this regard, the generated perturbation (error) vectors by Oscar will emerge in the approximate message $\hat{m}$ and Bob understands that there is a problem in $c_1$ and asks Alice to resend the random message $m$ again. As of now, Alice encrypts $m$ again as $c_2 = mG' + e_{P_2}$, $e_{P_1} \neq e_{P_2}$ and sends $c_2$ to Bob. This case is named message-resend condition [4]. Now, Oscar has two various encapsulated values $c_1$ and $c_2$ of the same random message $m$ and his objective is to retrieve $e_{P_1} = e_1 P$ and $e_{P_2} = e_2 P$ from $c_1 + c_2$. In the KEM-LDLC, because of utilizing the latin square LDLCs, the vectors $mG'$, $e_{P_1}$ and $e_{P_2}$ are rational and $c_1 + c_2$ does not essentially take on a specific form. In this situation, Oscar is unable to identify a message-resend condition with the assistance of $c_1 + c_2$. Furthermore, even if Oscar is able to signify the message-resend condition, he cannot ascertain the values of $e_{P_1}$ and $e_{P_2}$. As a result, the message-resend attack is unsuccessful.

## 5.5. The Embedding Attack

In the embedding attack [42], $n$ basis vectors of lattice $\mathcal{L}(G)$ and an encapsulated value $c$, for which the attacker desire for identifying a nearby lattice point in relation to it, are embedded in an $(n + 1)$-dimensional lattice $\mathcal{L}(G')$. In this case, by assuming that $G$ is the generator matrix of lattice $\mathcal{L}(G) \in \mathbb{Q}^n$, the generator matrix of lattice $\mathcal{L}(G') \in \mathbb{Q}^{n+1}$ is obtained as $G' = \begin{bmatrix} G & 0 \\ c & 1 \end{bmatrix}$, $G = \begin{bmatrix} g_1 \\ \vdots \\ g_n \end{bmatrix}$. At this point, the attacker employs a lattice reduction algorithm to identify the shortest nonzero vector within the $(n + 1)$-dimensional rational lattice $\mathcal{L}(G')$. The goal is to discover a vector whose first $n$ entries are the closest approximation to $c$. Experimental results indicate that the embedding attack, utilizing the lattice reduction algorithm to locate the shortest vector, is effective for lattice dimensions ranging from approximately $n = 110$ to $n = 120$. In the KEM-LDLC, the lattice aspect is sufficiently large, set at $n = 10^3$, which results in the failure of the embedding attack.

## 5.6. The Nguyen Attack

The Nguyen attack [43] is considered as a useful method to cut the lattice-based KEMs if the utilized perturbation error vector $e$ has a specific form, e.g., $e = \{\pm \beta\}^n$, where $\beta$ is a small integer unchanging. In this situation, by appending the integer vector $s = \{\beta\}^n$ to the encapsulated value $c$, the modular equation $c + s = mG' + e + s \equiv m_{2\beta} G'$ is obtained, where $m_{2\beta} = m \pmod{2\beta}$. By diminishing the vector $m_{2\beta} G'$ from the encapsulated value $c$, the vector $c - m_{2\beta} G' = (m - m_{2\beta})G' + e$ is obtained. After that, by dividing this equation by $2\beta$, the equation $(c - m_{2\beta} G')/2\beta = m'G' + e/2\beta$ is obtained, where $m' = (m - m_{2\beta})/2\beta$.

In this situation, since the rational point $(c - m_{2\beta}G')/2\beta$ is known, we can derive a simplified instance of the CVP that involves a significantly smaller perturbation error vector $e/2\beta \in \{\pm 1/2\}^n$. The length of the perturbation vector has now been reduced to $\sqrt{n/4}$, in contrast to the previous length of $\beta\sqrt{n}$. In reality, the challenge of decapsulating the encapsulated values is simplified to a more straightforward CVP instance in which the perturbation error vector consists of entries $1/2$. In the intended KEM-LDLC, the agitation error vector $e$ is a Gaussian random vector with a zero mean and variance $\sigma^2 < 1/(2\pi e)$. Consequently, the individual entries of the perturbation error vector cannot be determined exactly, which means that the modular equation $c + s = m_{2\beta}G'$ is not obtainable. Thus, the refined CVP example with a significantly smaller agitation error vector cannot be achieved, leading to the failure of the Nguyen attack in the proposed KEM-LDLC.

*5.7. The Round-Off Attack*

In the round-off attack [42], Oscar begins by multiplying the inverse of $G' = \text{HNF}(G)$ by the encapsulated value $c = mG' + e_P$ as $(G')^{-1}c = m + (G')^{-1}e_P$. He subsequently conducts an exhaustive search to determine the vector $d = (G')^{-1}e_P$. If the vector $d$ is successfully identified during this exhaustive search, the message $m$ can be accurately retrieved. As a result, the search space required to locate the precise vector $d$ must be adequately enormous to protect against round-off attacks. As it mentioned earlier, the HNF of a rational generator matrix $G$, represented as $G' = \text{HNF}(G)$, is also a rational matrix. Furthermore, the converse of the rational matrix $G'$ is also a rational matrix. Therefore, when multiplying the rational matrix $(G')^{-1}$ by the rational vector $e_P$, i.e., $d = (G')^{-1}e_P$ the resulting vector $d = (G')^{-1}e_P$ is likewise a logical vector. In this scenario, the quantity of rational vectors $d = (G')^{-1}e_P$ is boundless, making it impossible for an attacker to locate the vector $d$ through exhaustive search methods. As a result, this kind of attack is not feasible.

## 6. Conclusion

In this study, we explored the concepts of encoding and decoding algorithms for Latin square LDLCs to expand a KEM-LDLC that combines both code and lattice principles. We preserve the set of developing sequence $\mathcal{H}$ of used LDLC instead of the parity-check matrix $H$ to go down the SK magnitude. In addition, to reduce the PK size, we use the HNF of generator matrix $G = H^{-1}$, i.e., $G' = \text{HNF}(G)$, as the common key. In the intended KEM-LDLC, discovering the nearest lattice point $x$ to the encapsulated value is equivalent to solve the CVP. Also, retrieving the generator matrix $G$ from the common key $G'$ is identical to address the SBP. In addition, by using the Fujisaki-Okamoto conversion, we secure the proposed KEM-LDLC against the adaptive chosen ciphertext attack. The analysis manifests that the security and efficiency of the intended KEM-LDLC are influenced by several factors, encompassing lattice dimension, codeword length, perturbation (error) vector, and the LDLC decoding algorithm. Thus, to make a secure and efficient KEM-LDLC, it is essential to select and configure these factors in a manner that strikes a suitable balance between security and efficiency.